\documentclass[prb,reprint,showpacs,numerical,superscriptaddress,twocolumn,amsmath,amssymb]{revtex4-1}

\usepackage{graphicx}% Include figure files
\usepackage{dcolumn}% Align table columns on decimal point
\usepackage{bm}% bold math
\usepackage{amssymb}
\usepackage{color}
\usepackage[normalem]{ulem}

\begin{document}

\title{Out-of-plane and in-plane actuation effects on atomic-scale friction}

\author{O.~Y.~Fajardo}
\email{yovany@unizar.es}
\affiliation{Departamento de F\'{\i}sica de la Materia Condensada and
  Instituto de Ciencia de Materiales de Arag\'{o}n, CSIC-Universidad
  de Zaragoza, 50009 Zaragoza, Spain}

\author{E.~Gnecco}
\affiliation{Instituto Madrile\~no de Estudios Avanzados, IMDEA
  Nanociencia, 28049 Madrid, Spain}

\author{J.~J.~Mazo}
%\email{juanjo@unizar.es}
\affiliation{Departamento de F\'{\i}sica de la Materia Condensada and
  Instituto de Ciencia de Materiales de Arag\'{o}n, CSIC-Universidad
  de Zaragoza, 50009 Zaragoza, Spain}

\date{\today}

\begin{abstract}

The influence of out-of-plane and in-plane contact vibrations and
temperature on the friction force acting on a sharp tip elastically
pulled on a crystal surface is studied using a generalized
Prandtl{-}Tomlinson model. The average friction force is significantly lowered
in a frequency range determined by the `washboard' frequency of the
stick-slip motion and the viscous damping accompanying the tip
motion. An approximately linear relation between the actuation
amplitude and the effective corrugation of the surface potential is derived in
the case of in-plane actuation, extending a similar conclusion for
out-of-plane actuation.  Temperature causes an additional friction
reduction with a scaling relation in formal agreement with the
predictions of reaction rate theory in absence of contact
vibrations. In this case the actuation effects can be described by the
effective energy  or, more accurately, by
introducing an effective temperature.
\end{abstract}

%\keywords{}

\maketitle

% Introduction
\section{Introduction}

Sliding friction is a fundamental problem which is actively being
theoretically and experimentally studied at the micro- and
nano-scale~\cite{Bhushan2013,UrbakhNature2010,GneccoSpringer2007,UrbakhNature2004,PerssonKluwer1996,SingerKluwer1992}. Recent
experimental advances have provided a significant amount of new
information on the main characteristics of this phenomenon at small
scales. As a consequence, an important theoretical effort to
understand qualitatively and quantitatively the experimental outcomes
has been activated.\cite{Dong11,Vanossi13} In this frame, the problem
of reducing friction is of utmost importance.  Traditional lubricants
do not represent a valuable mean to achieve this goal in atomic-scale
contacts, since the hydrocarbon chains forming the lubricants cause
high viscosity when confined in nm-sized interstices~\cite{Hu98}.  An
efficient alternative is given by mechanical oscillations applied to
the sliding system~\cite{Dinelli97,Riedo03}.  This has
been shown in the atomic force microscopy (AFM) experiments in
  Refs.~\onlinecite{SocoliucScience2006} and~\onlinecite{Jeon06} in
which a dramatic reduction of friction was observed when the
oscillations were applied perpendicularly to the sliding plane
(\emph{out-of-plane}) and frequencies corresponding to mechanical
resonances of the probing tip coupled to the surface were chosen. In the same way Lantz \emph{et
  al.} could drive a silicon tip over a distance of several hundred
meters on a polymer surface without observing abrasive
wear~\cite{NN_Lantz09}.

Although the influence of out-of-plane oscillations on sliding
friction can be related to a periodic decrease of the energy
corrugation experienced by the slider~\cite{SocoliucScience2006},
systematic theoretical investigations of this effect at different
temperatures have not been reported so far, which is possibly due to
the lack of corresponding experimental results.  The same can be said
in the case of mechanical vibrations applied along the sliding
direction (\emph{in-plane}). Recent experiments~\cite{APL} have shown
a significant reduction of the average friction force at the atomic scale caused by
lateral vibrations applied to a nanotip elastically pulled on a flat
crystal surface. Here, we will try to model this type of {\em
  in-plane} actuation. Note that this experimental situation is
different from the substrate vibration problem studied numerically
in~\cite{Tshiprut05,Guerra08,Liu13}.

In this work we present detailed numerical simulations of sliding
friction on the atomic scale in the presence of out-of-plane and
in-plane ac fields. Results are obtained for different values of the
four main parameters of the system: frequency and amplitude of the ac
field, sliding velocity, and temperature.  Friction force is reduced
in well-defined frequency windows, and the effect is enhanced at
increasing temperature and actuation amplitude. An analytical formula
explaining the gradual transition from stick-slip to superlubricity at
$T=0$~K is also derived. Regarding the thermal behavior of the system,
we show how our numeric results scale following the law given by
reaction rate theory, commonly used to predict the temperature
dependence of atomic-scale friction in absence of mechanical
vibrations. Quantitative agreement is also obtained if appropriate
effective parameters are introduced.

% The Model
\section{The Prandtl-Tomlinson Model with ac Actuation}

In order to analyze ac-actuation effects on atomic-scale friction, we
start from the one-dimensional Prandtl-Tomlinson (PT) model including
thermal
effects~\cite{Prandtl28,GneccoPRL2000,SangPRL2001,SillPRL2003,ReguzzoniPNAS2009,OFajardoPRB2010,prl_jansen10,prb_muser12}:
\begin{equation}
\begin{split}
 m\dfrac{d^{2}x}{dt^{2}}+m\gamma \dfrac{dx}{dt}+\dfrac{\partial
   U(R,x)}{\partial x} = \xi(t),\\
   U(R,x)=V_\mathrm{el}(R,x)+V_\mathrm{int}(x).
\end{split}
\label{eqs1}
\end{equation} 
\noindent Here, a single particle of mass $m$, representing the AFM
tip apex, is driven along a periodic substrate, experiencing an
effective potential $U(R, x)$. This potential includes an elastic
contribution, $V_\mathrm{el}$, and a tip-substrate interaction term, $V_\mathrm{int}$.
%, see Fig.~\ref{Fig1}(a).
The first term describes the combined effect of the lateral
deformation of the elastic cantilever support (moving at a constant
velocity $v_s$, so its position is $R(t)=R_0+v_s t$), the deformation
of the tip apex, and of the sample surface:
$V_\mathrm{el}=k(v_st-x)^{2}/2$, where $k$ is an effective spring
constant~\cite{Carpick97,Lantz97}. In the first order approximation
the tip{-}surface interaction is $V_\mathrm{int}(x)=-U_0\cos\left(2\pi
x/a\right)$, where $a$ is the spatial periodicity of the surface
lattice. The random noise term $\xi (t)$ is related to the temperature
$T$ by the fluctuation{-}dissipation relation $\langle \xi(t) \xi (t
') \rangle = 2m\gamma k_{\rm B}T\delta(t-t') $, where $\gamma $ is a
microscopic friction coefficient describing the coupling with phonon
and possible electron oscillations in the substrate, and $ k_{\rm B} $
is the Boltzmann constant.

\begin{figure}[tb]
\centering
\includegraphics[scale=.45]{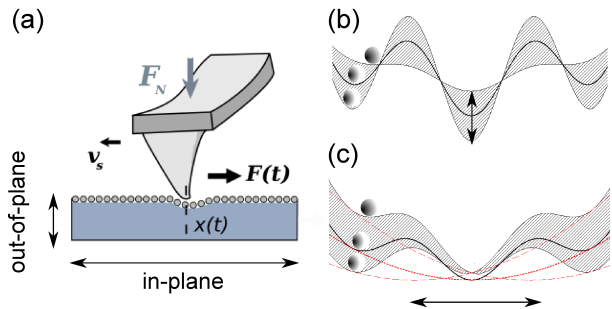}
\caption{(a) Sketch of a friction force microscopy experiment with
  actuation. A sharp tip is set into contact with a crystal surface
  under a given load $F_N$ and laterally pulled by a flexible
  cantilever at a constant velocity $v_s$.  The surface sample (or,
  equivalently, the free end of the cantilever) is shaken either in
  the normal or in the lateral direction.  Portraits of the potential
  energy $U(R,x)$ when the cantilever is set in motion are shown for
  (b) out-of-plane actuation and (c) in-plane actuation (in this case
  the red lines show the time variation of the elastic energy
  $V_\mathrm{el}$). The tip apex (represented by a gray ball) lies in
  a local minimum of the total potential $U(R,x)$.}
\label{model_fig}%Fig0
\end{figure}

Ac-actuation effects are taken into account in two different ways
(Fig. \ref{model_fig}).  Out-of-plane actuation acts perpendicularly
to the sliding plane. As discussed
in~\cite{SocoliucScience2006,Iizuka09}, this effect can
  be modeled assuming that the interaction potential oscillates as a
  function of time around the value $U_0$ (fixed by the normal load
  $F_N$) with the actuation frequency $f$ and a relative amplitude
  $\alpha$ [Fig. \ref{model_fig}(b)]:
\begin{equation}\label{outofplane}
V_\mathrm{int}(x,t)=-U_0\left(1+\alpha \cos 2\pi ft\right) \, \cos
\left( \frac{2\pi x}{a} \right).\end{equation} 

In the present work we
model \emph{in-plane} actuation by adding a shaking term to the
support position $R(t)$, so that the elastic energy reads~\cite{Igarashi07,Capozza11}
\begin{equation}
 V_\mathrm{el}(x,t)=\dfrac{k}{2} \left(  v_{s} t + \beta a\ \sin  2\pi
   f t  -x \right)^{2} .
\label{pot2}
\end{equation}
In Eq.~(\ref{pot2}) the parameter $\beta$ quantifies the relative
amplitude of the oscillations.  The variation of the total potential
$U(R,x,t)$ during an oscillation is sketched in Fig. \ref{model_fig}(c).

In the following, we are interested in estimating the mean value
$\langle F \rangle$ of the instantaneous lateral force $F(t)= k
[R(t)-x(t)]$, as measured by the AFM, at different model parameters,
e.g. the amplitude and frequency of the ac-actuation term, scan
velocity and temperature.  In the framework of the PT model, the
quantity $\langle F \rangle$ can be identified with the kinetic
friction force acting on the tip.  To this end we have integrated the
equations of motion of the system, Eq.~(\ref{eqs1}), and averaged the
results over many periods of the ac field. Non-dimensional equations
are introduced by measuring energy in units of $U_0$, length in units
of the lattice spacing ($\widetilde{x}=2\pi x/a$) and time in units of
the resonance frequency of the point mass in the wells of the
sinusoidal potential ($\tau=\omega_p t$ with
$\omega_p=2\pi\sqrt{U_0/ma^2}$). Then $\widetilde{\gamma}=\gamma /
\omega_p$, $\widetilde{k}=1/\eta=k a^2 /(4\pi^2U_0)$ and
$\widetilde{v_s}=v_s\sqrt{m /U_0}$ are the scaled damping, spring
constant and velocity respectively.

For the parameter values we have chosen $m= 2.0\times 10^{-12}$\,kg,
$a=0.246$\,nm, and $k=3.73$\,N/m.  For $v_s$ we have used 25~nm/s,
except for the results shown in Fig.~\ref{Fig2} where two values,
$v_s=$~25 and 250 nm/s, have been chosen. With respect to the
potential amplitude $U_0=0.25$~eV. This set of values correspond to
typical parameters used in experiments and theoretical works on the
topic~\cite{SangPRL2001,OFajardoPRB2010, OFajardoJPCM2011}.  For our
choice of parameter values $\omega_p/2\pi=0.6$\,MHz, and $\eta=7.0$.

\section{Actuation effects at zero temperature}

\begin{figure}[tb]
\centering
\includegraphics[scale=1.1]{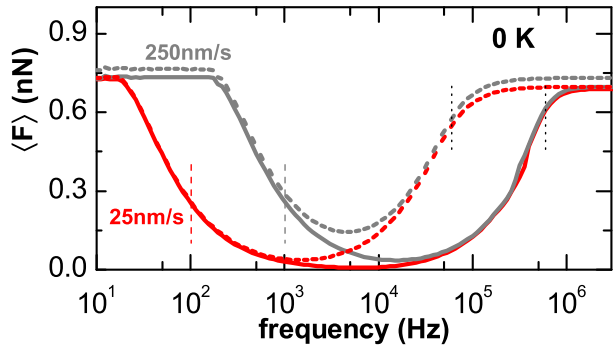}
\caption{(Color online) Damping and velocity effects on the friction
  force in presence of out-of-plane actuation. We show the mean friction force
  as a function of the actuation frequency $f$ and for two velocity
  values, $v_s=25$ (red line) and $250$ nm/s (gray line). Continuous
  and dashed line correspond to the nondimensional damping values
  $\widetilde{\gamma}=1$ and $10$, respectively. The vertical dashed
  lines indicate the frequencies $f_{\rm wb}=v_s/a$ and
  $\omega_p/2\pi\widetilde{\gamma}$ associated with both velocities
  and dampings. These lines define the limits for which the reduction
  of the friction becomes effective.}
\label{Fig2}
\end{figure}

We first review results for ac-actuation without thermal effects.
Figure~\ref{Fig2} presents the most important ones, where we have
chosen to show the effect of out-of-plane actuation at $\alpha=0.9$, a
large amplitude value. Results for smaller $\alpha$ and in-plane
actuation are qualitatively similar and are summarized in
Sec. \ref{sec_thermal}. The most important finding is the existence of
a wide medium frequency range ($f \sim$ a few kHz) where the friction
force is importantly reduced, and almost suppressed for intense enough
actuation.  We notice that the lower bound of this
reduced friction zone is determined by the support velocity $v_s$, and
the upper bound by the effective damping $\gamma$. Indeed, friction
reduction is observed for actuation frequencies $f$ such that
\begin{equation}\label{frequency_range}
v_s/a \leq f \leq \omega_p/(2\pi \widetilde{\gamma}).\end{equation}

The support velocity $v_s$ defines a characteristic frequency of the
system, the washboard frequency $f_{\rm wb}=v_s/a$ associated to the
time needed to advance one period of the substrate potential. Thus,
for frequency values $f\ll f_{\rm wb}$, actuation is not
effective. This is not the case when $f$ approaches $f_{\rm wb}$ as
seen in Fig.~\ref{Fig2} (where $f_{\rm wb}$ is marked by vertical
dashed lines for each velocity value). In addition, damping and
inertial effects do not play an important role in this low frequency
range.

At high frequency we can consider the system as an overdamped forced
oscillator. The tip is unable to follow the external field for $2\pi
\gamma f/\omega_p^2 \gg 1$, which corresponds to $f \gg
0.6/\widetilde{\gamma}$ MHz with our choice of values. In this high
frequency regime the support velocity $v_s$ does not play an important
role.

\subsection{Dependence of friction on the actuation amplitude}\label{sec_linear}

\begin{figure}[t]
\centering
\includegraphics[scale=0.17]{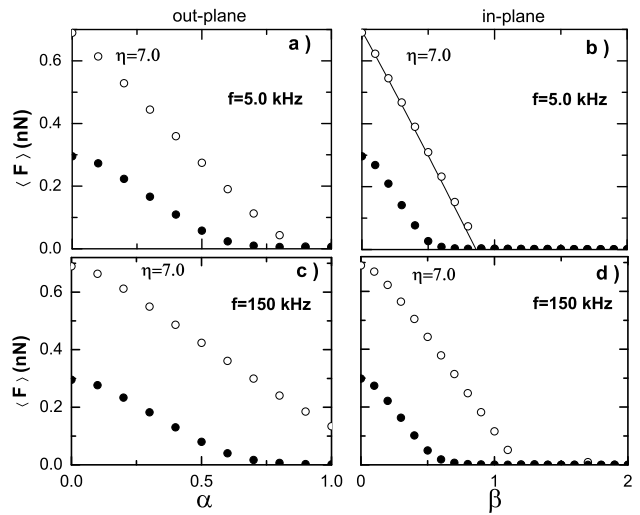}
\caption{Friction force $\langle F \rangle$ for different values of
  the actuation amplitude for the frequencies values $f =5$ kHz and
  $f=150$ kHz.  (a) and (c) correspond to the out-of-plane actuation
  case; (b) and (d) to in-plane actuation. In all cases $T=0$ K for
  open circles and $300$ K for solid circles. Full line in figure (b)
  stands for the theoretical prediction of Eq.~(\ref{final_formula})
  up to the third term on the RHS.}
\label{Fig3}
\end{figure}

One of the most important questions is to determine how the overall
friction depends on the actuation field amplitude. Such issue is
studied in Fig.~\ref{Fig3}. Here the mean friction force
$\left<F\right>$ is represented as a function of the amplitude
oscillation for two frequency values with and without thermal
effects. Figure~\ref{Fig3} shows the decrease of the force
$\left<F\right>$ as a function of $\alpha$ or $\beta$ down to a
`frictionless' superlubricity regime.  As shown in Fig.~\ref{Fig3},
the magnitude of the reduction effect varies almost linearly with the
amplitude of the actuation field. We note that this effect can be
reproduced by defining an effective amplitude barrier $U_0^{\rm eff}$,
and thus an effective PT parameter $\eta_{\rm eff}$, corresponding to
the $\eta$ value which gives the same friction force of the original
system with an applied actuation field of intensity $\alpha$ (or
$\beta$). In other words, $\langle F \rangle (\eta,\alpha)=\langle F
\rangle (\eta^{\rm eff},\alpha=0)$ (and similarly for $\beta$). Such
$\eta^{\rm eff}$ can be computed for each value of $\alpha$ or $\beta$
and $f$.

\begin{figure}[tb]
\centering
\includegraphics[scale=0.18]{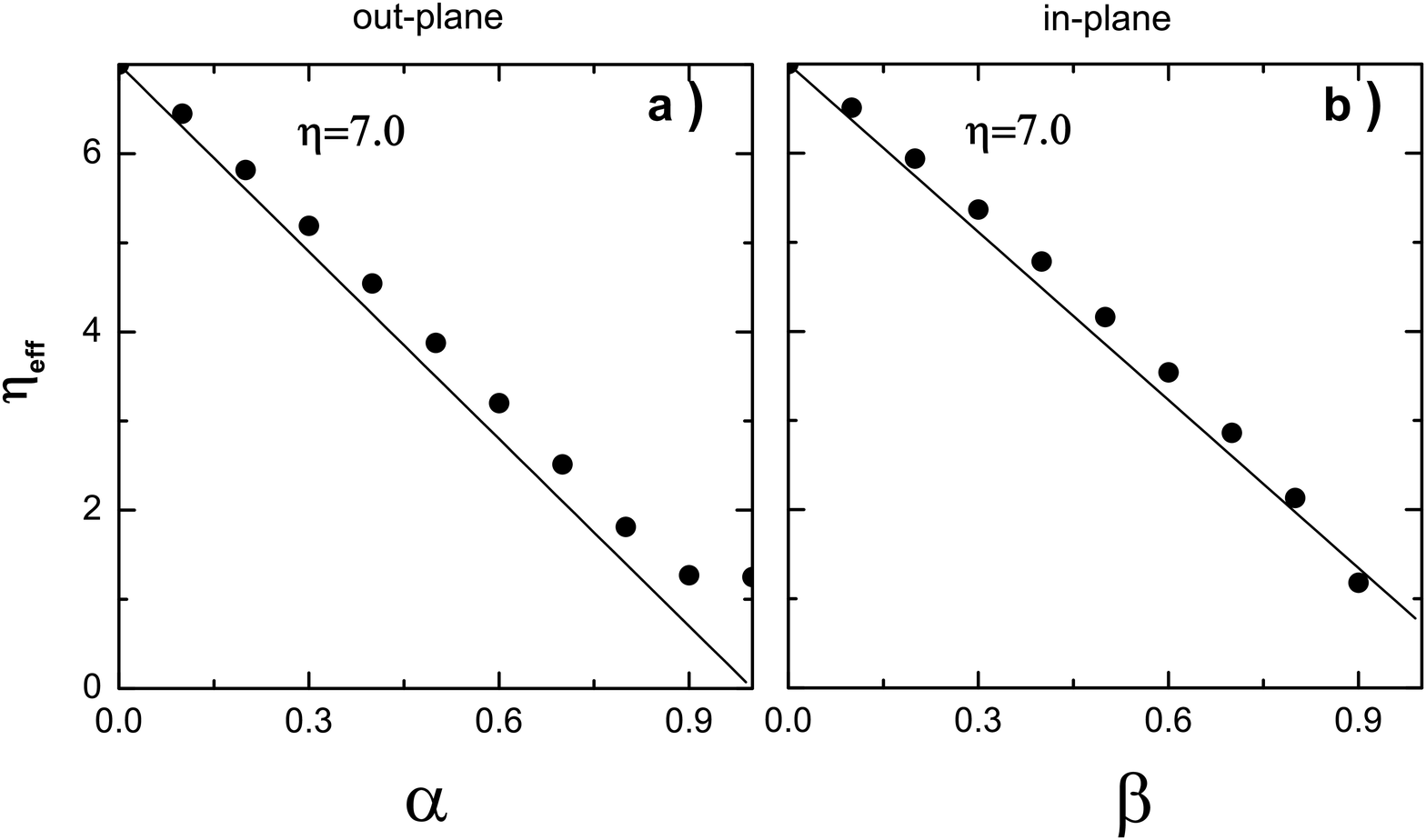}
\caption{(Color online). Theoretical and numerical comparison of the
  effective PT parameter for different values of the actuation
  amplitude at $f=5$ kHz and $T=0$.}
\label{Fig4}%Fig0
\end{figure} 

At $T=0$ analytical expressions for the dependence of
$\eta_\mathrm{eff}$ with the actuation amplitude can be derived. These
expressions are effective at the frequency values corresponding to the
minima of the friction force vs. frequency curves. In case of
out-of-plane actuation, Socoliuc \emph{et
  al.}~\cite{SocoliucScience2006} observed that
\begin{equation}\label{eta_alpha}
\eta_\mathrm{eff}(\alpha)\simeq
\eta(1-\alpha),
\end{equation}
considering the minimum value of the energy barrier for slippage which
is reached during an oscillation.  Similarly, in case of
in-plane-actuation, we have found that
\begin{equation}\label{eta_beta}
\eta_\mathrm{eff}(\beta) \simeq \eta-2\pi\beta.
\end{equation}
Equation (\ref{eta_beta}) is obtained from the theoretical analysis
presented in the Appendix. There, assuming that the slip event occurs
at the maximum elongation of the tip-support spring, we are able to
predict the dependence of the friction force with the actuation field
amplitude for the case of {\em in-plane} actuation and zero
temperature, Eq.~(\ref{final_formula}). Such expression is derived
considering the first nonzero terms in the series expansion in $\beta$
of the `take-off' and `landing' positions of the tip slipping across
the surface lattice and the corresponding variation of the total
energy $U$. The predicted result is plotted in figure~\ref{Fig3}(b)
which corresponds to a frequency for which friction is strongly
reduced in the system. Furthermore, Fig.~\ref{Fig4}
  shows that, if one wants to keep to a linear dependence on $\eta$, a
  good approximation is obtained using $\eta_{\rm eff} =7.0-6.44\alpha$
  and $7.0-6.0\beta$. The numeric coefficients appearing in these
  expressions are only slightly different from those in
  Eqs. (\ref{eta_alpha}) and (\ref{eta_beta}).

\begin{figure}[tb]
\centering
\includegraphics[scale=0.35]{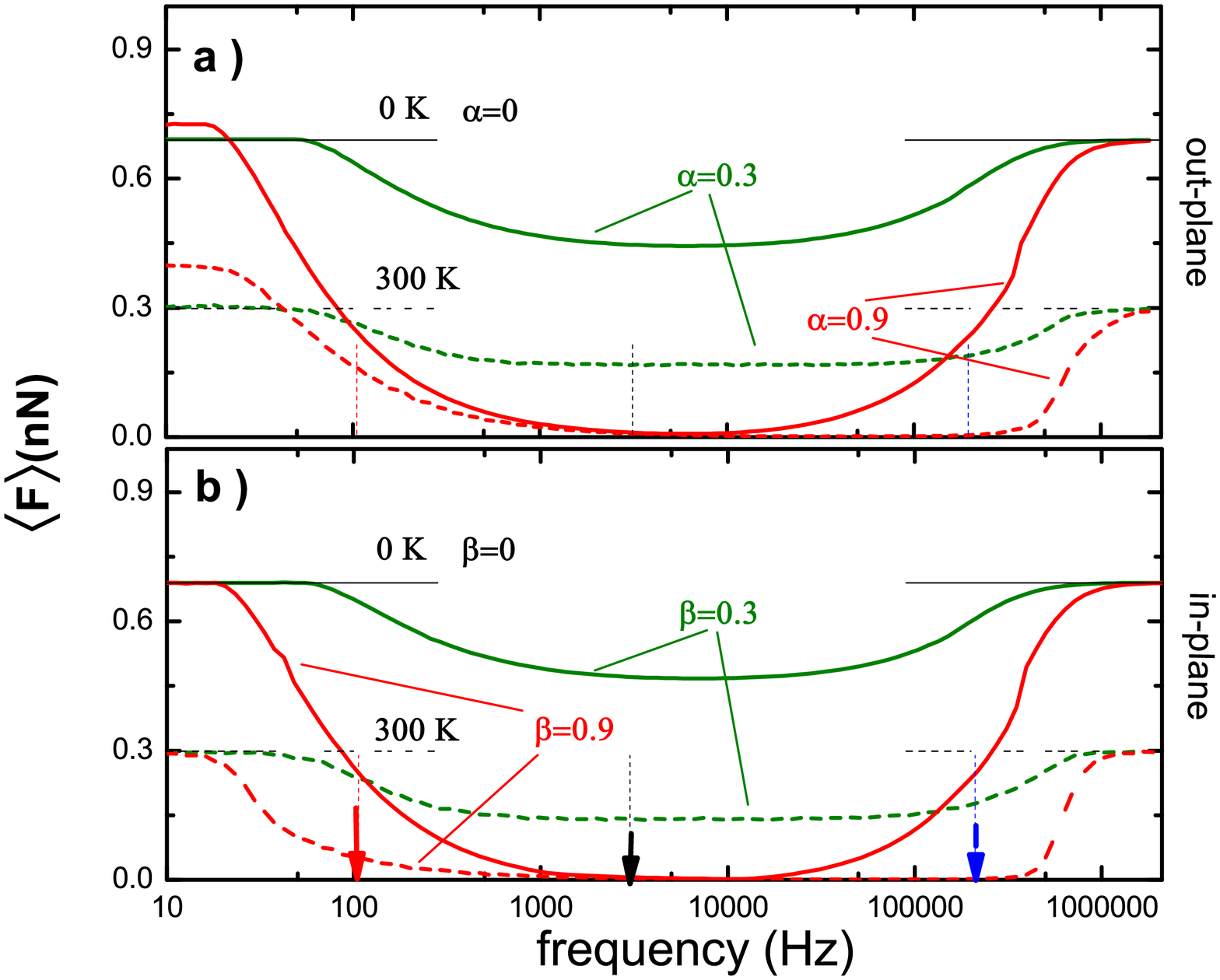}
\caption{(Color online). Friction force as a function of the actuation
  frequency for $T=0$ K (continuous lines) and $T=300$ K (dashed
  lines). $v_s=25$ nm/s and $\widetilde{\gamma}=1$.  (a) Out-of-plane
  actuation curves for $\alpha =$ 0.3 (upper green lines) and 0.9
  (lower red lines). (b) Results for in-plane actuation for $\beta=$
  0.3 (upper green lines) and 0.9 (lower red lines). Thin vertical
  lines correspond to $v_s/a=101$ Hz, and two specific frequency
  values used in our study, $f=5$ kHz and 150 kHz.}
\label{Fig5}%Fig0
\end{figure} 

\section{Thermal effects}\label{sec_thermal}

\begin{figure}[tb]
\centering
\includegraphics[scale=0.35]{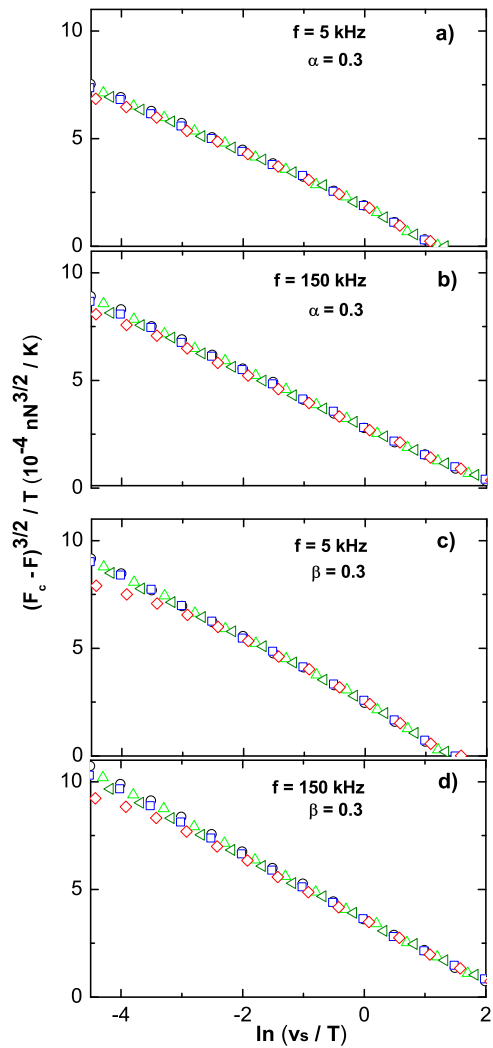}
\caption{(Color online). Scaling behavior of the friction force in presence of
  out-of-plane actuation with $\alpha =$0.3 and (a) $f=5$ kHz (b) $f=150$
  kHz and in presence of in-plane actuation with $\beta=$0.3 and (c) $f=5$ kHz, (d) and $f=150$ kHz. Circles, triangles, squares, lateral triangles and diamonds
  correspond to the temperatures $T= 150$, $200$, $250$,
  $293$ and $373$ K, respectively.}
\label{Fig6}%Fig0
\end{figure}

\begin{figure}[]
\centering
\includegraphics[scale=0.17]{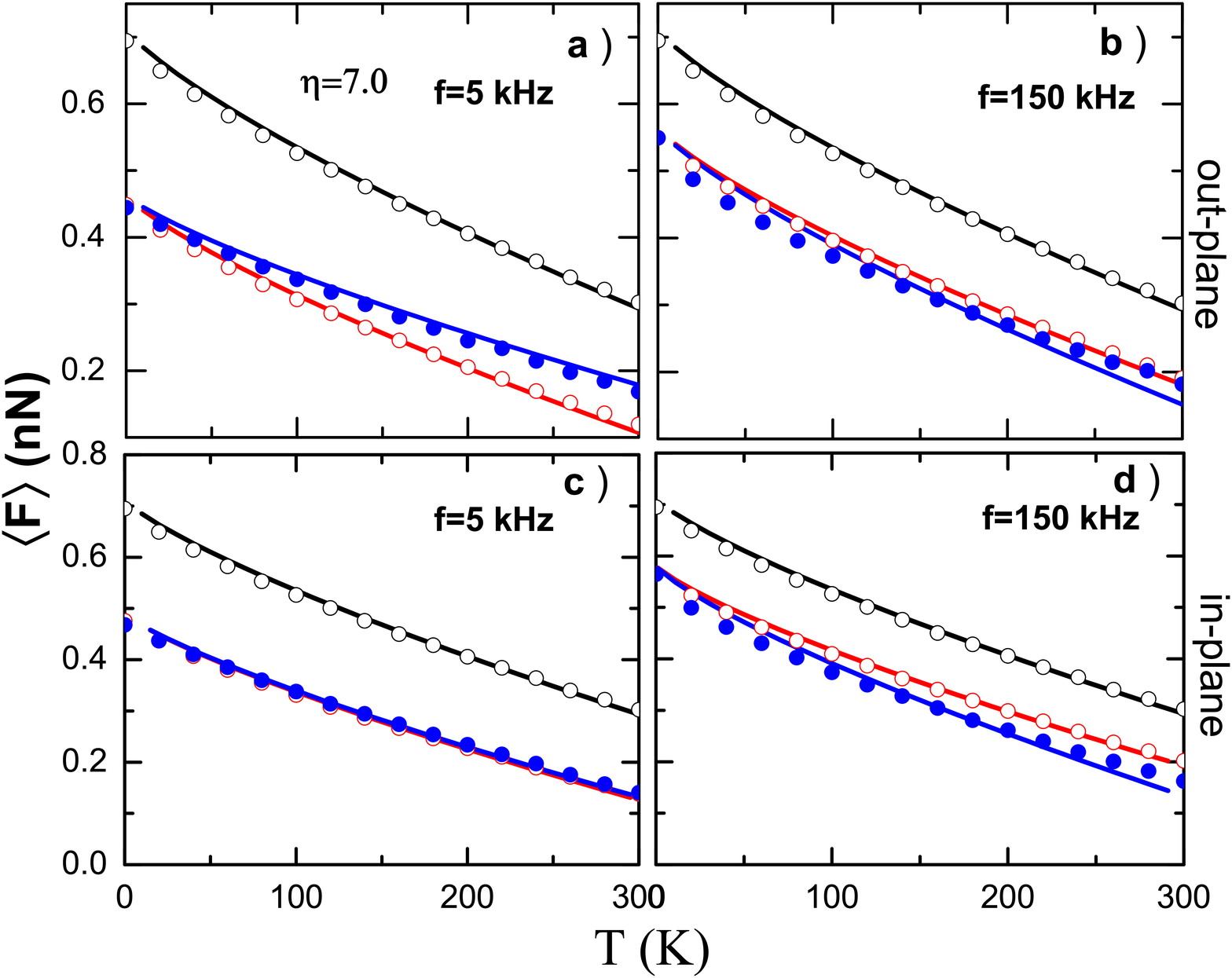}
\caption{(Color online). Friction force at different temperatures with
  and without ac-actuation (either out-of-plane or in-plane):
  simulations without actuation (open black circles) for $\eta=7.0$
  and fits from Sang's theory (black curves); simulations without
  actuation using an effective energy corrugation (open red circles)
  and fits from Sang's theory (red curves); simulation with actuation
  (solid blue circles) and fits using an effective energy corrugation
  and effective temperature $T_{\rm eff}$ (blue curves)
  with $\theta=$0.82, 1.1, 0.98 and 1.2 from (a) to (d).}
\label{Fig7}%Fig0
\end{figure}

It is well know that friction at the atomic scale can be reduced by
thermal vibrations\cite{SangPRL2001}. At finite temperature the tip
slippage is anticipated, which decreases the average friction
force. This is also the case when an actuation force is
applied. Figures~\ref{Fig3} to~\ref{Fig7} show the dependence of the
mean friction force on the actuation frequency $f$, the actuation
field amplitude $\alpha$ or $\beta$, and the temperature $T$. These
figures reveal the existence of a frequency range where the actuation
effectively reduces friction, an almost linear dependence of the
friction force on the actuation amplitude, and also describe the
influence of thermal vibrations on the friction force.

In particular, Fig.~\ref{Fig3} compares the dependence of the friction
with the actuation amplitude at $T=0$ K and $T=300$ K. It can be seen
that an almost linear decrease holds also in the presence of thermal
effects.

The change in the friction vs. actuation frequency curves when thermal
effects are introduced is shown in Fig.~\ref{Fig5}. As expected, an
important reduction in the friction force is observed. This reduction
adds to that caused by the ac-actuation on the tip. We also observe
that the frequency boundaries of the region of effective friction
reduction are not modified at finite temperature.

\subsection{Scaling behavior}

To account for the temperature dependence of the mean friction force
in the absence of ac-actuation, the following law has been
proposed~\cite{Prandtl28,SangPRL2001}:
\begin{equation}
\langle F \rangle=F_c-B \, T^{2/3} \, [{\ln}(C \, T / v_s)]^{2/3},
\label{eqSang}
\end{equation}
where $F_c$, $B$ and $C$ are constants defined by the system
parameters. Equation (\ref{eqSang}) can be obtained from the Kramers
expression for the escape rate of a particle out of a metastable well,
evaluated in the overdamped limit.

The case of ac actuation force is much more complex. In fact, no
simple expression for the escape rate has been derived in the presence
of ac fields~\cite{xx,yy,zz}. As a consequence, a simple relation for
the friction force reduction is also missing.  Nevertheless, as
substantiated by the numerical simulation results detailed below, we
observe that the formula (\ref{eqSang}), with different values of the
parameters $F_c$, $B$ and $C$, is still valid when both ac actuation
and thermal effects are included.
This can be seen in Fig.~\ref{Fig6}, where results for both out-of-plane and in-plane actuation and two different frequency values  are shown. Here, the same scaling relation predicted in absence of actuation fields,
\[\frac{(F_c-F)^{3/2}}{T}\propto \log\frac{v_s}{T}+\mathrm{const.},\]
is found to hold. In Fig.~\ref{Fig6} both actuation amplitudes
$\alpha$ and $\beta$ were set equal to 0.3, a moderate value. This
scaling only breaks down for large enough actuation fields, when $F$
goes to zero and the superlubricity regime is reached (see
Fig.~\ref{Fig3}).
 
\begin{figure*}[tb]
\centering
\includegraphics[scale=0.4]{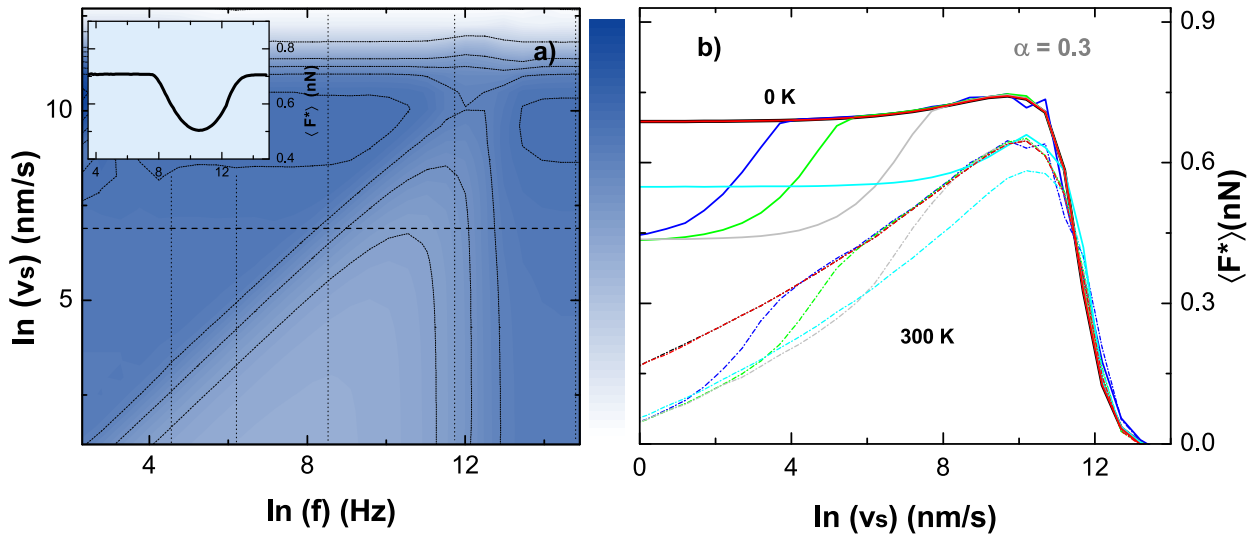}
\caption{(Color online). (a) Iso-surface of the friction force as a
  function of the logarithm of the tip velocity and the excitation
  frequency in case of out-of-plane actuation without thermal
  effects. The inset shows a profile of the friction force along the
  dashed line marked on the iso-surface. (b) The friction force as a
  function of the tip velocity for different frequencies values: $f=0$
  (black), $100$ (blue), $500$ (green), $5\times 10^{3}$ (light gray),
  $150\times 10^3$ (cyan) and $3\times 10^6$ (red) Hz. Continuous
  curves correspond to $T=0$ K and dashed curves to $T=300$ K.  In all
  cases $\alpha=$0.3 and $\widetilde{\gamma}=1$.}
\label{Fig8}%Fig0
\end{figure*}

\subsection{Effective temperature}

Our major finding is that the ac-actuation effect can be described by
introducing new effective parameters of the system. This approach is
found to be valid in a wide range of actuation parameters values and
in the stick-slip regime of the system. In Sec. III. A we introduced
an effective PT parameter, $\eta_{\rm eff}$, which at zero temperature
accounts for the friction reduction in the system when ac fields are
applied. Now, guided by the scaling relations we have found, and by
the existing theories, we will extend our discussion to the
understanding of the friction versus temperature curves.

Figure \ref{Fig7} summarizes
our conclusion. Open black circles describe the temperature dependence
of the friction force in the absence of ac-actuation. We see that the
friction force is well fitted by Eq.~(\ref{eqSang}) of the Sang
\emph{et al.} theory~\cite{SangPRL2001} (black lines in
Fig.~\ref{Fig7}).  The solid blue circles in Fig.~\ref{Fig7} show the
temperature dependence of the friction force in presence of actuation.
These points are approximately fitted by the red curves, corresponding
to Eq.~(\ref{eqSang}) with $\eta$ replaced by the parameter $\eta_{\rm
  eff}$ discussed in the previous section.

However, as expected from the obtained scaling dependence, a much
better agreement is obtained if, in addition, an effective temperature
$T_{\rm eff}=\theta T$ is introduced (where $\theta$ is close to
1). This is showed by the blue curves, corresponding to the equation
\begin{equation}
\langle F \rangle=F_c-B \, T_{\rm eff}^{2/3} \, [{\ln}(C \, T_{\rm eff} / v_s)]^{2/3},
\label{eqF}
\end{equation}
where the parameters $F_c$, $B$ and $C$ are calculated in the same way
as in Eq.~(\ref{eqSang}), using the obtained effective value for the
energy of the corrugation potential. Only at high values of the
actuation amplitude strong nonlinear effects destroy the scaling
relationship between the real temperature $T$ of the system and the
effective temperature $T_\mathrm{eff}$.

\subsection{Force-velocity curves}

To conclude, we present in Fig.~\ref{Fig8} additional results on the
friction force versus velocity dependence with ac actuation and
thermal effects. We have to remark that in this case we plot the
reduced friction force $\langle F^{*} \rangle =\langle F \rangle
-\gamma m v_s$ in order to suppress the viscous drag contribution
which dominates at high velocities. Damping has been fixed to
$\widetilde{\gamma}=1$. Inset of Fig.~\ref{Fig8}(a) shows a typical
force-frequency curve, similar to those showed in
Fig.~\ref{Fig2}(a). Figure~\ref{Fig8}(a) shows a log-log surface plot
at $T=0$ of the reduced friction force as a function of dragging
velocity $v_s$ and actuation frequency $f$. As previously discussed
the low friction region narrows as the support velocity
increases. This is due to the increasing value of $f_{\rm wb}$ which
gives the onset of the small friction zone, whereas the high frequency
boundary, which basically depends on the $\omega_p/\widetilde{\gamma}$
ratio, remains constant.

Figure~\ref{Fig8}(b) shows the friction versus velocity curves at
different frequency values and temperatures $T=0$ and 300 K.  Such
curves can be easily understood with the help of Fig.~\ref{Fig8}(a)
where we have marked the chosen frequency values. There, thermal
effects on the velocity dependence of the friction force are
explicitely plotted. Such effects can be understood with the help of
Eq.~(\ref{eqF}) after the introduction of the effective parameters.

\section{Conclusions}

We have made a detailed numerical analysis of in-plane and
out-of-plane ac actuation effects on the average friction
  force of a system at the atomic scale, in the framework of an
one-dimensional Prandtl-Tomlinson model including thermal effects.
Our work shows that in-plane and out-of-plane actuation give pretty
similar results.  This can be qualitatively understood from
Fig.~\ref{model_fig} (b) and (c). In both cases one observes a
periodic reduction of the energy barrier to be overcome by the
tip. While this is not surprising in the case of out-of-plane
actuation, the same effect for in-plane actuation is less obvious and
quite remarkable. 

Actuation forces are effective for friction reduction in an
intermediate range of frequencies. This range is limited by the
washboard frequency of the system, associated to the motion at a given
velocity over a periodic potential, and by the frequency value at
which the relaxation time of the system becomes comparable to the
actuation period so that the system becomes unable to follow the
dynamics of the ac force. The first frequency is essentially given by
the drag velocity of the support, $v_s$; the second one is determined
by the effective damping $\gamma$ of the system.

Having seen that the ac actuation effect can be understood in terms of
a reduction of the effective barrier of the system which accounts for
a reduction of the PT parameter, $\eta$, thermal effects cause an
additional friction reduction which is well reproduced by the Sang
expression of Eq.~(\ref{eqSang}). A better agreement is obtained
introducing an effective temperature weakly different from the real
one.

The friction reduction caused by in-plane oscillations and the scaling
relations predicted by our work remain to be fully tested experimentally. It
is interesting to observe that, in absence of ac actuation, Jansen
\emph{et al.} found a good agreement with the thermally activated PT
model at any temperature between 100 and 300 K on
graphite\cite{jansen10}. However, they also observed a decrease of
friction with decreasing $T$ below 130 K on NaCl\cite{barel11}, which
was related to the formation of multi-asperity contacts. Superimposing
lateral vibrations at different frequencies while sliding may help to
shed light on this intriguing effect.

Other actuation schemes has been recently theoretically
studied. Interesting results using simulations of many-atoms
confinated between two plates, one sliding, the other vertically
vibrating also showed friction suppression at suitable vibration
frequency and amplitude~\cite{Capozza09}.

\begin{acknowledgments}

O.~Y. Fajardo and J.~J. Mazo acknowledge financial support from
Spanish MICINN through Project No. FIS2011-25167, cofinanced by FEDER
funds. 
E. Gnecco acknowledges financial support from Spanish MINECO
through Project. No. MAT2012-34487.
 O.~Y. Fajardo acknowledges financial support from FPU grant by
Ministerio de Ciencia e Inovaci\'on of Spain.

\end{acknowledgments}

\begin{appendix}

\section{Analytical expressions for atomic stick-slip with in-plane actuation}

Here we will present an analytical approach to $\langle F \rangle
(\eta,\beta)$, and discuss how Eq.~(\ref{eta_beta}) in the main text
can be derived as a first approximation for the not trivial dependence
of the average friction force on the parameter values $\eta=4\pi^2
U_0/(ka^2)$ and $\beta$.  If the driving velocity is sufficiently low,
the tip position $x$ at a given time $t$ is given by the equilibrium
condition $\partial U/\partial x=0$. In absence of vibrations,
\begin{equation}\label{extra0}
U(x,t)=-\eta\cos x+\frac{(v_st-x)^2}{2}\end{equation}
(in nondimensional units), and the previous condition becomes
\begin{equation}\label{extra1}
\eta \sin x+x-vt=0.\end{equation}

When the spring is pulled along the scan direction the tip slightly
moves forwards.  When the tip reaches the critical position $x_c$
defined by $\partial^2 U/\partial x^2=0$, the equilibrium becomes
unstable and the tip jumps.  The critical position is given by
\begin{equation}\label{extra2}
\eta \cos x_c+1=0.
\end{equation}
Combining (\ref{extra1}) and (\ref{extra2}) we get for the critical time $t_c$:
\[ v t_c=\sqrt{\eta^2-1}+\arccos (-1/\eta)\equiv f(\eta). \]

When in-plane vibrations are switched on, the jump will be
anticipated. In order to calculate the critical time in this case, we
have to replace $vt$ with $vt+\beta$ in Eq.~(\ref{extra1}).  This
substitution makes sense provided that the spring elongation does not
change significantly during the period $1/f$ of the oscillations,
which is the case if the condition (4) in the text is satisfied.

After jumping the tip ends into the next minimum $x'_c$ of the total
potential $U$. If $\eta \gg 1$ the coordinates $x_c$ and $x'_c$ of the
take-off and landing points are approximately equal to $\pi/2$ and
$5\pi/2$ respectively.

As a next step, assuming that $\eta$ is large enough, we expand the
function $f(\eta)$ introducing the variable $\nu=1/\sqrt{\eta}$:
\[ f(\eta)=\frac{1}{\nu^2}+\frac{\pi}{2}+... \]
The take-off and landing positions are then given by
\begin{equation}\label{before}
x_c=  \frac{\pi}{2}-2\sqrt{\pi\beta}\nu+\nu^2-\frac{\pi^{3/2}\beta^{3/2}}{3}\nu^3+\frac{2\pi\beta}{3}\nu^4-...
\end{equation}
and
\begin{eqnarray}\label{after}
x'_c= && \frac{5\pi}{2}-2\sqrt{\pi(1+\beta)}\nu+\nu^2+  \nonumber \\
&& -\frac{\pi^{3/2}}{3}(1+\beta)^{3/2}\nu^3+\frac{2\pi}{3}(1+\beta)\nu^4-...
\end{eqnarray}
respectively.

At each jump the energy amount $\Delta U = U(x_c) - U(x'_c)$ is
released from the contact region in the form of phononic or electronic
excitations.  The mean friction force, which we identify as the
lateral force averaged over several lattice constants, can be simply
determined from this quantity as $\left<F\right>=\Delta U/2\pi$, in
nondimensional units.  Substituting
the expressions (\ref{before}) and (\ref{after}) in (\ref{extra0}) we
finally get
\begin{widetext}
\begin{equation}\label{final_formula}\left<F\right>(\eta,\beta)=
\eta-\pi(1+2\beta)+\frac{4}{3}\sqrt{\frac{\pi}{\eta}}\left[(1+\beta)^{3/2}-\beta^{3/2}\right]-\frac{1}{2\eta}
+\frac{2}{15}\left(\frac{\pi}{\eta}\right)^{3/2}\left[(1+\beta)^{5/2}-\beta^{5/2}\right]-...,
\end{equation}
\end{widetext}
Note that the equations (\ref{before}-\ref{final_formula}) generalize
the relations (15-17) in Ref. \onlinecite{prb_Gnecco12} when $\beta\ne
0$. Previous equations are dimensionless. In standard units, the right
hand sides of the equations defining the critical positions and the
friction force are multiplied by the factors $(a/2\pi)$ and
$(ka/2\pi)$ respectively.

In particular, when $\beta=0$:
\[
\left<F\right>(\eta,0)=
  \eta-\pi+\frac{4}{3}\sqrt{\frac{\pi}{\eta}}-\frac{1}{2\eta}+\frac{2}{15}\left(\frac{\pi}{\eta}\right)^{3/2}-... \]
At the first order in $\beta$, we can thus say that
\[ \left<F\right>(\eta,\beta)\simeq \left<F\right>(\eta-2\pi\beta,0),\]
i.e. the effect of lateral vibrations on the support can be approximately taken into
account replacing the parameter $\eta$ with
$\eta_\mathrm{eff}=\eta-2\pi\beta$.

\end{appendix}

\end{document}